# New clustering method to decrease probability of failure nodes and increasing the lifetime in WSNs

Shahram Babaie
Department of Computer Engineering
PhD students, Islamic Azad University,
Olom VA Tahghighat Branch,
Tehran, Iran
.

Ahmad Khadem Zade
Department of Computer Engineering
Iran Telecommunication Research Center
Tehran, Iran
.

Ali Hosseinalipour
Department of Computer Engineering
Islamic Azad University- Tabriz Branch
Tabriz Iran
.

*Abstract*—Clustering in wireless sensor networks is one of the crucial methods for increasing of network lifetime. There are many algorithms for clustering. One of the important cluster based algorithm in wireless sensor networks is LEACH algorithm. In this paper we proposed a new clustering method for increasing of network lifetime. We distribute several sensors with a high-energy for managing the cluster head and to decrease their responsibilities in network. The performance of the proposed algorithm via computer simulation was evaluated and compared with other clustering algorithms. The simulation results show the high performance of the proposed clustering algorithm.

I. INTRODUCTION

Recent improvements in integrated circuits (IC) have fostered the emergence of a new generation of tiny, called Sensors. That from economical aspect they are commodious and also they are used in non military (for instance environmental managing: temperature, pressure, tremor, etc)

To consider having group of limitations such as battery life time, calculating and memory significantly they have been predicted as non recyclable and also they live until their powers fade away. So power is something rare for systems like sensor. During a special mission the correct consumption for sensors lifetime should be managed knowingly. The power of sensor can not support more than far connection. Therefore to transmit they need the architecture of multi sectional. A useful way to decrease the system lifetime is to divide them to diverse clusters [2]. Parts of a cluster-based network sensor are base stations and sensors. In this method sensors relay the data flow by head clusters. The central station always stays far from where sensors are expanded. In this manner saving the consumption energy and awareness of that to communicate with central station has various methods. Two methods of routing in articles have been proposed [5, 6]. These methods because of their route detecting and finding optimum steps in relation with central station have head load. In addition, they will have extra load on nodes that are located around central station, so most of the traffic will be from them.

To avoid these overheads and unbalanced consumption of energy some high-energy nodes called "Gateways" are deployed in the network [2]. These sensors are used as head clusters due to decrease the failure probability of head clusters. And this increases the lifetime of the network. But since this method takes a lot of expenditure so in this article we just use these sensors as manager for a number of head clusters. In this manner each one becomes gatewayamong each head cluster. This method decreases both networks lifetime and failure probability.

In the second part, the architecture of two networks and the relevant tasks will be explained. In the third part, the proposed protocol has been explained and in the fourth part the results of simulating and tests evaluation can be seen. The last part involves conclusion of the article and discussing about pattern of future researches.

II. RELATED WORKS

System architecture for clustered sensor networks has been shown in figure 1. There just two sorts of nodes, cluster joint sensors and head cluster with tolerance of energy shortcoming. Joint sensors and homogeneous head clusters with a same identity have been assumed as similar. All the connections are wireless. The connection of joint nodes with the main station is possible only with head cluster. For sending information schedule we use TDMA (Time-Division Multiple Access) protocol.

During starting the process a unique ID, primary energy and TDMA scheduling are attributed for all the sensors and gateways. We suppose that the entire node are aware from others place by the GPS. In the beginning all of the connective bridges are assumed in connection area. as the energy consumption of GPS is high , it is On at the beginning of the clustering and on the other states it is in the sleep mode. Connection scheduling among connective bridges first appears with head cluster when it establishes.

The central station always stays far from where the sensors are expanded. in this order, maintaining the consumption energy and being aware of that in relation with central station have different methods: such as LEACH (Low-energy Adaptive Clustering Hierarchy) [1] and SEP (Stable Election Protocol) [7] and also two other routing method have been explained in articles [5, 6].these methods due to detecting the path and finding the optimum steps in relation with command node have head load. In addition having extra load on nodes, which is located around central station, most of the traffics will be because of them. To avoid this head loads and unstable energy consumption some of nodes have been expanded through the networks by a high-energy that called gateway [2].these sensors act as gateway among clusters and central stations. And mage the entire network in cluster. Each sensor with a high-energy belongs just to one cluster. And the connection with the central station just takes place through cluster Gateway. In this method failure probability decreases and networks lifetime increases.

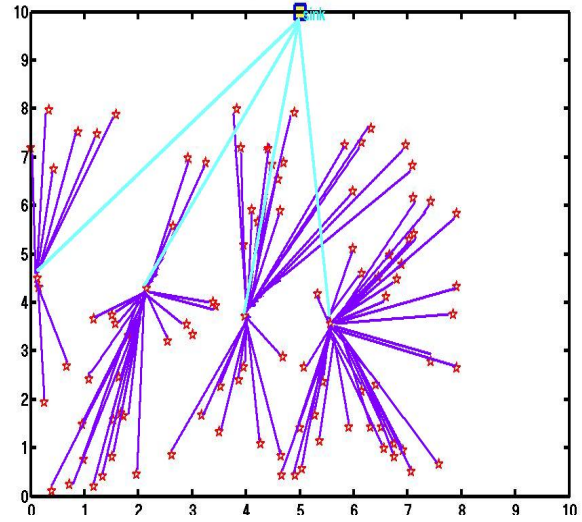

Figure 2: The LEACH Network model

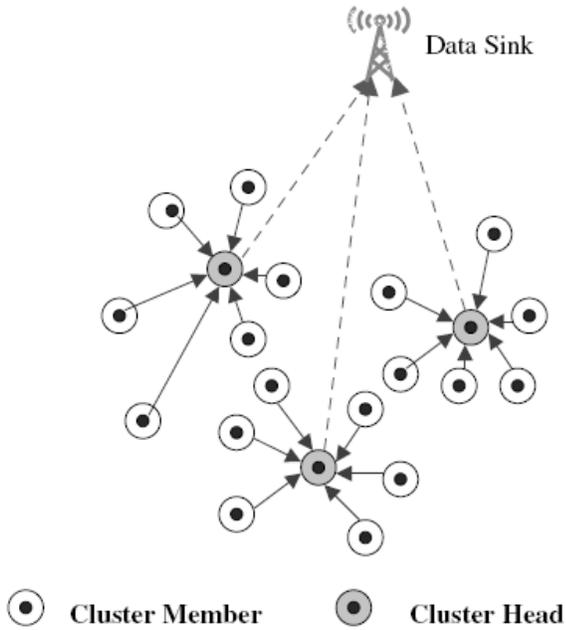

Figure 1: network style with clustering

### A. LEACH Protocol

LEACH protocol is hierarchical routing algorithm that can organize nodes into clusters collections. Each cluster controlled by cluster head. Cluster head has several duties. First one is gathering data from member cluster and accumulates them. Second one is directly sending accumulation data to sink. Used model in LEACH shows in Figure 2. Third one is scheduling based of Time-Division Multiple Access (TDMA). In that, each node in cluster related to it's time slot could send collection data [1].

Cluster head announce time slot by uses of distribution property to all members of cluster. Main operations of LEACH are classify in two separate phase that shows in figure 3 [1]. First phase or initialization phase has two process; clustering and cluster head determining. Second phase mean steady-state, that this phase concentrate to gathering, accumulation and transmit data to sink.

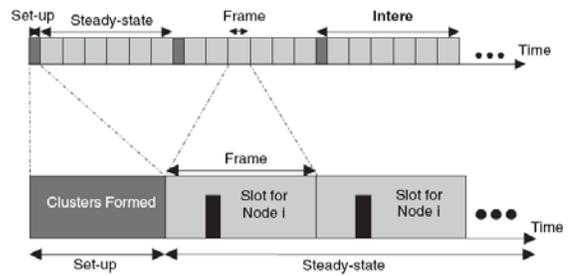

Figure 3: The LEACH protocol Phase.[1]

First phase as a compared with second one less overhead impose to protocol. In initialization phase, at first in choose of cluster head step, randomly allocate number between zero and one for each node and then compared with cluster head threshold. A node is chosen for cluster head if its number is less than threshold. Threshold of cluster head shows in relation 1.

$$T(n) = \begin{cases} \dfrac{P}{1 - P\left(r \bmod \left(1/p\right)\right)} & n \in G \\ 0 & n \notin G \end{cases}$$

Relation 1: The way of cluster head selection relation.[1]

T (n): Threshold
P: node selection probability in every period
G: the set of unselected nodes
r: number of current period (started from one)

A node as a cluster head, announces new roles to the other nodes. With this signal other nodes based on straight of received signal decide to be membership of which cluster. In every cluster, created time slot by cluster head based on TDMA, distributed between cluster that contain visual time slot for each member of cluster. Cluster head use Code-Division Multiple Access (CDMA) technique too. With completion of initialization phase, steady-state phase start. In this phase, nodes in determining time slot gathering data and sending to the cluster head node. Meanwhile gathering of data is periodically.

*B. SEP Protocol*

SEP protocol was improved of LEACH protocol. Main aim of it was used heterogeneous sensor in wireless sensor networks. This protocol have operation like LEACH but with this difference that, in SEP protocol sensors have two different level of energy. Therefore sensors are not homogeneous. In this protocol with suppose of some sensors have high energy therefore probability of these sensors as cluster head will increased. But in SEP and LEACH, cluster heads aren't choose base of energy level and their position. This is main problem of these methods, so their operations are static.

### III. PROPOSED METHOD

These methods because of their route detecting and finding optimum steps in relation with central station have head load. In addition, they will have extra load on nodes that are located around central station, so most of the traffic will be from them.

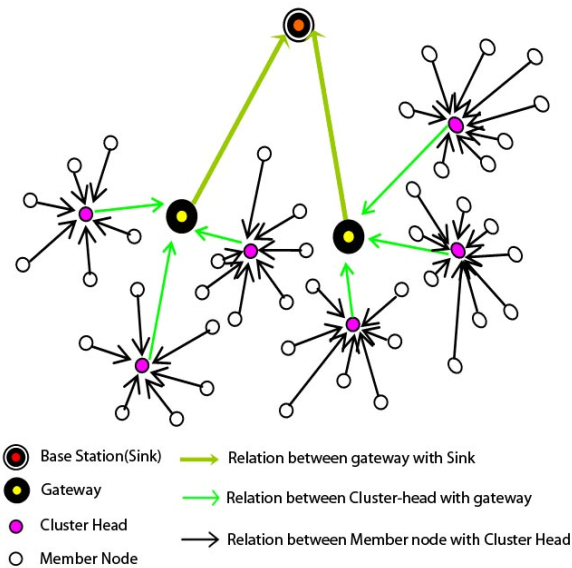

Figure 4: Multi-gateway clustered sensor network

To avoid this extra load and unstable consumption of energy some of the nodes have been expanded with a high-energy called Gateway [2] sensors are used as head clusters due to decrease the failure probability of head clusters. And this increases the lifetime of the network but since this method takes a lot of expenditure so in this article we just use these sensors as manager for a number of head clusters.

To do so, we expand some of nodes according to lifetime, space and number of exist sensors in network. While clustering we don't need this work node. We can cluster the network with the algorithms like SEP, LEACH and TEEN (Threshold-sensitive Energy-Efficient sensor Network Protocol).afterward the clustering is done, each head cluster sends a signal to these sensors. And with these signals the sensors specify which cluster is appropriate to manage. And with the hypothesis of network they choose some of the cluster to in order to manage them. And each closer is being managed just by one of these sensors. After establishing the network the role of these sensors as gateways between head clusters and central stations, by the hypothesis network chooses some of clusters to manage. And each cluster is being controlled by just one of the sensors. After establishing the network, the sensors have the role of a gateway between central stations and head clusters. To be attentive that head clusters to transmit to central stations and data assembling and calculating in protocol consume a great deal of energy. All the responsibility of head cluster is given over to joint cluster sensors or Gateway. Then after receiving data from its joint nodes without any calculating delivers them to gateway. And its gateway that transmits them to base station after doing necessary works and calculations. This method can be used in two ways. One that we spread high-energy sensors beside other sensors. And another practical way is to put them between root station and head clusters. In both aspects both network lifetime increases and extra load eliminates from head clusters and also failure probability decreases.

That other cluster heads don't have connection with Sink station. And this connection is accomplished via Gateway and these nodes with high-energy contain the rule of Gateway. And these Gateways to lifetime termination managing same cluster heads. But similar to LEACH algorithm in any time period the cluster head is changing. When the cluster node is changing, the cluster head tell to gateway via a signal. This protocol is resumed to end of lifetime.

### IV. SIMULATION RESULTS

We stimulated a wireless sensor network in a 100*100 space and with an equal distribution of 100 sensors randomly by using MATLAB software. In this simulation the central node at the end of area with the co ordinations has been put. And we spread 4 sensors with high power in network. The primary energy of typical sensors is 0.5 J and sensors with high-energy are 1.0 J. we adjust the execution of the simulation for 1000 cycle and also consumption energy is evaluated based on table number 1.

TABLE 1: USED RADIO CHARACTERISTICS IN OUR SIMULATIONS

| Operation | Energy Dissipated |
|---|---|
| Transmitter/Receiver Electronics | Eelec=50nJ/bit |
| Data Aggregation | EDA=5nJ/bit/signal |
| Transmit Amplifier if dmaxtoBS ≤ d0 | $\epsilon fs$=10pJ/bit/m2 |
| Transmit Amplifier if dmaxtoBS ≥ d0 | emp=0.0013pJ/bit/m4 |

The results of simulation show that new method in comparison with LEACH and SEP acts better and also increases the networks lifetime significantly.

We test this protocol and LEACH and SEP with different sensors (50,100,200,300,400,500) and as seen in figure 5 the results show that the new method is better than exist methods. And the lifetime of the network is more than the same lifetime in LEACH and SEP. both LEACH and SEP die with 100 sensors when they see the first sensor and live for another 200 time. While in the proposed protocol after observing the first died sensor that itself observes later than LEACH and then lives for another 300 times.

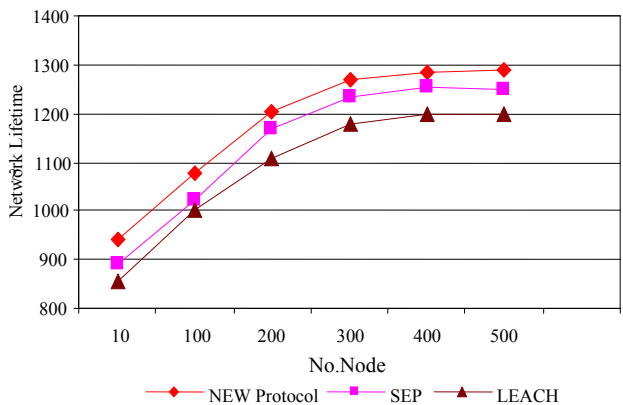

Figure 5: Comparing proposed algorithm with others

## V. CONCLUSION AND FUTURE WORKS

The node of Gateway with a high-energy through the sensors is used as a central manager is just a step away from the central station. Ultimately after simulating we found out that proposed protocol plays an indispensable role in increasing network lifetime and could have been increased the lifetime in comparison with SEP and LEACH.

In this article it is supposed that sensor nodes and gateways are fixed and motionless. On the other program we will research the mobile gateways.